\documentclass{PoS}
\usepackage{amsmath}                
\usepackage{amsfonts}               
\usepackage{amssymb}                
\usepackage{epsfig}                 

\newcommand{\cm}{{~\rm cm}}
\newcommand{\km}{{~\rm km}}
\newcommand{\s}{{~\rm s}}

\newcommand{\erg}{{~\rm erg}}
\newcommand{\yr}{{~\rm yr}}

\newcommand{\pc}{{~\rm pc}}
\newcommand{\kpc}{{~\rm kpc}}

\newcommand{\AU}{{~\rm AU}}

\title{Evolution of the Jet-Feedback Mechanism (JFM)}
\ShortTitle{Evolution of the of the JFM}

\author{\speaker{Noam Soker}\thanks{This research was not supported by any grant}\\
        Department of Physics, Technion , Haifa, Israel\\
        E-mail: \email{soker@physics.technion.ac.il}}

\abstract{
I list eight types of astrophysical objects where jets, and more particularly the jet feedback mechanism (JFM), might operate, and discuss cases where an object evolves from one type to another while the JFM continues to operate.
In four of these classes of objects jets are known to play significant, or even crucial, roles: in cooling flows, during galaxy formation, in young stellar objects (YSO), and in planetary nebulae.
In core collapse supernovae (CCSNe), in the common envelope evolution (CEE), in the grazing envelope evolution (GEE), and in intermediate-luminosity optical transients (ILOTs) the suggestion that a JFM takes place is still controversial.
\emph{I call for a refresh thinking and more detail studies of the possibility that jets play a large role in exploding massive stars and in the CEE.}
I also present a new speculative scenario where the first active galactic nuclei (AGN) were preceded by a JFM that operated during the life time of the supermassive young object (SMYO)  progenitor of the AGN. A short and energetic phase of CCSN took place between the SMYO and the AGN phases.
I term this scenario of young object to supernova to AGN (YOSA) that includes a JFM along all stages, the YOSA-JFM scenario.
\emph{I speculate that in the YOSA-JFM scenario, the JFM that might have operated during the phase of the SMYO started to establish the correlations between the mass of the super-massive black hole (SMBH) and some properties of the stellar component of galaxies before the formation of the SMBH. } }

\FullConference{Frontier Research in Astrophysics - II\\
		 23-28 May 2016\\
		 Mondello (Palermo), Italy}

\begin{document}

\section{Introduction}
\label{sec:intro}

Large varieties of astrophysical systems contain a compact object, orders of magnitudes smaller than the system, that in principle can accrete gas from the system and liberate huge amounts of gravitational energy. In many cases the most efficient way by which the gravitational energy can be tapped to the system is by two opposite jets launched by the accreting compact object.
The back influence of the jets on the system, that is basically the reservoir of gas feeding the compact object, might lead to a negative feedback mechanism. This is the jet-feedback mechanism (JFM).
In a recent paper (\cite{Soker2016}) I reviewed the JFM in different types of systems, as summarized in Table \ref{table:compare}. In \cite{Soker2016} I discuss the important ingredients of the JFM, the similarities and differences between the different systems, the roles of the JFM in the different systems, and much more. I will not repeat the discussions from that review. In the present paper I present the main new points from my talk and discussions during the meeting, which are basically the possible evolution of the JFM from one type of object to another.
\begin{table*}
\tiny
    \begin{tabular}{|l||l|l|l|l|l|l|l|l|}
     \hline
Property &Clusters & Galaxy-F & CCSNe & PNe & CEE&GEE&ILOTs&YSOs  \\
 \hline
  \hline
Energy (erg)&$10^{60}$&$10^{59}$&$10^{51}$&$10^{44}$&$10^{44-48}$& $10^{44-48}$ & $10^{46-49}$ &       $10^{43-46}$\\
             \hline
Mass $(M_\odot)$& $10^{12}$ & $10^{11}$  & $10$  &$1$  & $1$  & $1$ & $1-10$ & $1-10^3$\\
              \hline
Size; $R_{\rm res}$&$100 \kpc$&$10 \kpc$&$10^9 \cm$&$0.1 \pc$& $10^{1-2}R_\odot$& $10^2R_\odot$ & $10^2R_\odot$ & $10^{3-5} \AU$\\
              \hline
$-\Phi_{\rm res}$ &$(1)^1$&$(0.3)^2$&$(10)^2$&$(0.03)^2$ &$(0.03)^2$ & $(0.03)^2$&$(0.1)^2$ & $(0.001)^2$ \\
          \hline
Time  & $10^{7-8} \yr$ & $10^{7-8} \yr$ & $1-3 \s$ & $10^{1-2} \yr$ & $1-100\yr$ & $10-100\yr$ & $0.1-10\yr$ & $10^{2-5}\yr$  \\
              \hline
$T_{\rm bubble}$(K) & $10^{9-10}$ & $10^{9-10}$ & $10^{10}$ & $10^6 $ & $10^{7}$  & $10^{7}$ & $10^{7}$ & $10^{3}$\\
 \hline
$T_{\rm ambiant}$(K)& $10^{7-8}$ & $10^{6-7}$ &few$\times 10^9$ & $10^4$ & $10^{5-6}$& $10^{5}$ & $10^4$ & $100$\\
      \hline
ComObj & SMBH & SMBH & NS/BH  & MS/WD       & MS/WD/NS & MS & MS & MS     \\
 mass $(M_\odot)$ & $10^{8-10}$& $10^{6-9}$ & $1-50$ & $1$ & $1$ & $1$ & $1-10$ & $1-10$ \\
$R_a$(cm) & $10^{13-16}$ & $10^{11-14}$ & $10^{6}/10^{7}$ & $10^{11}/10^{9}$  & $10^{11}/10^{9}/10^{6}$ & $10^{11}$ & $10^{11-12}$ & $10^{11-12}$ \\

$-\Phi_a$ & $c^2$ & $c^2$ & $(0.1-1)c^2$ & $(0.5)^2/(5)^2$  & $(0.5)^2/(5)^2/$  & $(0.5)^2$ & $(1)^2$& $(0.5)^2$ \\
& &  & & & $(100)^2$  &  & & \\
\hline
 $\Phi_a/\Phi_{\rm res}$&$10^5$&$10^6$&100&100&100&100&3-100&$10^5$\\
\hline
Jets' main     &Heating&Expelling&Exploding&Shaping&Removing   &Removing&Reducing &Expelling\\
$\qquad$ effect&the ICM&     gas & the star& the PN&part of the&envelope&accretion&gas;\\
               &       &         &         &       &envelope   &        &         &Turbulence\\
    \hline
Role of & Maintain   &$M_{\rm BH}-\sigma$&Explosion &Not&Might     &Ensures &Not &Slowing  \\
the JFM & ICM        &correlation  &energy $\approx$&much&limit    &outer   &Much&star  \\
        & temperature&             &binding         &    &accretion&envelope&    &formation \\
        &            &             & energy         &    &rate     &removal &    &\\
    \hline
Observation&X-ray   &Massive&(Axi-   &Bipolar& ~      &Bipolar&Radiation;&Bipolar\\
           &bubbles;&outflow&symmetry)&PNs    & ~      &remnant&bipolar   & outflows\\
           &cold gas&       &         &       & ~      &       &remnant   &  \\
    \hline
Fizzle &Cooling    & Rapid & BH        & &Core-    & Forming & &More           \\
outcome&catastrophe& SMBH+stars& formation;& &secondary& a common& &gas forms  \\
       &           & growth& GRB       & &merger   & envelope& &stars  \\
    \hline
Importance&Crucial in&Very &Contestant &Jets common;&Might occur &Crucial&Not    &Common;  \\
of jets/JFM&all CFs   &common; &with neutrino&the JFM does&in some cases;&&crucial&not   \\
          &          &not crucial&mechanisms&not operate &not crucial   &&       &crucial    \\
    \hline
Status of  &In       &In       &In fierce&In       &Not in   &Newly   &Not in   &In           \\
jets/JFM in&consensus&consensus&debate   &consensus&consensus&proposed&consensus&consensus\\
community  &         &         &         &         &         &        &         &           \\
\hline
    \end{tabular}
    \caption{Systems discussed in this paper where feedback and/or shaping by jets take place.
    The different listed values are typical and to an order or magnitude (or two even) accuracy only.
    Typical energy: Energy in one jet episode.
    Typical Mass, Size: of the relevant ambient gas. Typical time: the duration of the jets activity episode.
    In the row of observations, in parenthesis are expected observations.
    \newline
    Abbreviations and acronyms:
$R_a$ is the typical radius of the accreting object, and $\Phi_a$ is the magnitude of the gravitational potential on its surface in terms of the light speed $c$ or in units of $(1000 \km \s^{-1})^2$.
$R_{\rm res}$ stands for the typical radius of the reservoir of gas for accretion onto the compact object (the size of the system), and $\Phi_{\rm res}$ is the specific energy required to expel the reservoir (energy per unit mass) from the system in units of $(1000 \km \s^{-1})^2$.
  \textbf{Galaxy-F}: galaxy formation; \textbf{PNe}: Planetary nebulae; \textbf{CCSNe}: core collapse supernovae; \textbf{CF}: cooling flow; \textbf{ICM}: Intra-cluster medium; \textbf{CEE}: common envelope evolution; \textbf{GEE}: grazing envelope evolution;     \textbf{ILOTs}: intermediate-luminosity optical transients; \textbf{YSOs}: young stellar objects; \textbf{BH}: black hole; \textbf{SMBH}: super-massive BH; \textbf{NS}: neutron star; \textbf{WD}: white dwarf; \textbf{MS}: Main sequence star; \textbf{ComObj}: The compact object that accretes mass and launches the jets; $M_{\rm BH}-\sigma$ stands for the correlation of the SMBH mass with the stellar velocity dispersion. Taken from \cite{Soker2016}.
   }
 \label{table:compare}
\end{table*}

\section{Evolution }
\label{sec:properties}

I discuss the evolutionary relations between different systems. For the operation of the JFM in each type of system the reader should consult the review paper (\cite{Soker2016} and references therein).

\subsection{GEE~$\rightarrow$~ILOT~$\rightarrow$~CEE}

In the common envelope evolution (CEE) the JFM is not a necessary process. Nonetheless, I argue that jets facilitate the removal of the common envelope (CE), and that the jets operate through a JFM.
When the jets are very efficient in removing the envelope, in particular the envelope outside the orbit of the secondary star, the secondary star is not immersed in the envelope of the giant star, but it rather grazes the giant star. This is the grazing envelope evolution (GEE).
If the jets interact with previously ejected gas, a large fraction of the kinetic energy can be channelled to radiation, leading to an outburst that is generally called intermediate luminosity optical transients (ILOTs; subgroups of ILOTs and other names are red novae, red transients, intermediate-luminous red transients, SN impostors, major outbursts of luminous blue variables (LBV)).
The newly proposed GEE (\cite{Soker2015}) cannot take place without the JFM. The operation of the ILOT, like the CEE, does not require the JFM, but recently we (\cite{KashiSoker2016}) proposed that the JFM might operate in some ILOTs to some degree.

A giant star can enter a GEE with a compact companion (a main sequence star, a white dwarf, or a neutron star). Then the accretion process releases energy and jets. The jets might collide with ambient gas close to the binary system and lead to a bright outburst, an ILOT. The jets then might operate in a negative feedback mechanism not only in the dynamical evolution itself, the GEE, but also in regulating the outburst. If the jets do not stay efficient, and/or the spiraling-in process is too rapid, the secondary star enters the envelope of the giant star, and a CEE commences.  We have the evolution \textbf{GEE~$\rightarrow$~ILOT~$\rightarrow$~CEE}.
If the jets regain their relative efficiency, we might have the following evolution while the JFM continues to be active \textbf{CEE~$\rightarrow$~GEE}, and the GEE might be accompanied by an ILOT.
\subsection{Galaxy formation~$\rightarrow$~Cluster}
During the formation process of young galaxies the main roles of the JFM is probably to expel baryonic mass from the galaxy and to established the correlation between the mass of the super massive black hole (SMBH) and some properties of stars in the galaxy (see table). Such a JFM operates for a limited duration. A short-live cooling flow might take place during galaxy formation, but it will cease to exist after the supply of gas ends. If however the galaxy sits in a deep potential well more gas can be accumulated to form an extended hot medium. This can be the case for a large elliptical galaxy, the central galaxy in groups of galaxies, or the central galaxy in clusters of galaxies. In such a case there might be a continuous transition of a JFM operating during galaxy formation to a JFM that regulates the cooling flow for billions of years.

\subsection{Supermassive YSO~$\rightarrow$~CCSN~$\rightarrow$~AGN}

This evolutionary sequence is the most speculative scenario presented here. In this scenario the JFM operates continuously from a young (stellar) object (YO; not necessarily a star) through a CCSN, and then in a newly born AGN. The proposed YO~$\rightarrow$~SN~$\rightarrow$~AGN (acronym: YOSA) scenario is based on that the formation of SMBH at high red-shifts might be accounted for by the core collapse of a supermassive star (e.g., see a recent review by \cite{LatifFerrara2016}).
The properties of the YOSA-JFM scenario are summarized in Table \ref{table:YOSA}
\begin{table*}
\small
    \begin{tabular}{|l||l|l|l|}
     \hline
Phases /                   &Supermassive Young     & Supernova & Active galactic  \\
Property                   &(stellar) object (SMYO)& (CCSN)    & nucleus (AGN)  \\
  \hline
Duration                      & $10^6 \yr$ & 1~day      & $ > 10^6-10^7 \yr$       \\
Accreting body      &  Supermassive star   & new BH     &  SMBH      \\
Mass $(M_\odot)$              & $10^4-10^6$& $10^4-10^6$& $>10^5$     \\
Size $(R_\odot)$              &  $10^3$    & $1$        & $> 1 $       \\
$v_{\rm jet}$ $(\km \s^{-1})$  & $10^4$     & $0.1c-c$   & $c$      \\
$M_{\rm jets}$ $(M_\odot)$    & $10^3-10^5$& $10^3-10^5$& $>10^4$   \\
$E_{\rm jets}$ (erg)          & $10^{55}-10^{57}$& $10^{56}-10^{58}$ &  $> 10^{57}$   \\
\hline
Role of &Early $M_{\rm BH}-\sigma$ & Causes an  & $M_{\rm BH}-\sigma$\\
the JFM &  correlation             & explosion & correlation    \\
\hline

\hline
    \end{tabular}
    \caption{The phases of the YOSA-JFM scenario, where the JFM operates continuously during the life of a supermassive young object (SMYO), then during the formation of a BH in a collapse where the JFM drives a CCSN, and ending with the formation of an AGN. Properties listed are the duration of the phase, the accreting object, its mass, its size, the velocity of the jets, the mass carried by the jets, and the energy carried by the jets. In the AGN phase time can be short or can be up to the age of the universe. Quantities are given to an order of magnitude. }
 \label{table:YOSA}
\end{table*}

The evolution starts with the formation of a supermassive, $M_{\rm YO} \approx 10^4-10^6 M_\odot$, young object. The young object accretes mass from an accretion disk at a rate of $\gtrsim 0.1 M_\odot \yr^{-1}$, and live for about $1-4 \times 10^6 \yr$ \cite{Begelman2010}. For a mass of $M_{\rm YSO} \approx 10^6 M_\odot$ and a radius of $\approx 10^3 R_\odot$ (e.g., \cite{Begelman2010}), the escape velocity from this young (stellar) object is $\approx 2 \times 10^4 \km \s^{-1}$. The jets (or a collimated disk wind, which I refer to as jets as well) from the accretion disk around such an object can release an energy of $\approx 10^{55}-10^{57} \erg$, equivalent to $\approx 10^4-10^6$ CCSNe. The jets affect the cloud from which the supermassive star was formed in a feedback mechanism, and beyond. Due to the very large energy of the jets from the YSO, they influence a large fraction of the ISM, heating it and expelling part of it from the young galaxy.
I actually speculate that the JFM that might determine the correlation between the mass of the SMBH and some properties of stars in the galaxy \emph{ might start operating before the AGN is formed.}
I note that in a competing scenario where the SMBH is formed from a nuclear stellar cluster, SNe that are formed by the stars in the star cluster can also remove gas from the cloud. But I do not study this process here.

The collapse of the core of the SMYO is likely to take place while the accretion process from the ISM onto it did not terminate yet. After the center of the SMYO collapses, it forms a SMBH. Due to the large specific (per unit mass) angular momentum, an accretion disk is formed around the newly born SMBH. Jets are most likely to explode part of the star, mainly along the polar directions, but material stays bound near the equatorial plane and continues to feed the newly born SMBH. The average accretion rate during the explosion is the stellar mass divided by the free fall time from its surface, $\approx 1~$day. This gives $\dot M_{CCSN} \approx 10 M_\odot \s^{-1}$, not much different from that in regular CCSNe.
After the explosion, the accretion process continues on a very long time scale. With a mass of $M_{\rm SMBH} \gtrsim 10^4 M_\odot$ and a long lasting accretion, the object has turn to an AGN, that affects the ISM via a JFM.

Begelman et al. (2006) \cite{Begelmanetal2006} discuss the feedback of energy that is released by the mass accretion on to the newly born SMBH. They, however, do not mention jets, but only radiation, and, therefore, conclude that the liberated energy is trapped inside the quasi-star. The launching of jets changes the flow structure envision by \cite{Begelmanetal2006}. In particular, the jets can penetrate out of the quasi-star and influence the ISM. The JFM operates differently than feedback that is powered by radiation.

I note the following regarding this speculative YOSA-JFM scenario.
\newline
(1) There are three consecutive phases of JFM: A YSO-JFM, a CCSN-JFM, and a JFM in galaxy formation. In some case the system might evolve further to a JFM operating in a cooling flow.
\newline
(2) Neutrino cooling is not important and cannot account for such an explosion. Although the accretion rate is as in regular CCSNe, the radius of the newly formed SMBH is more than three orders of magnitude larger than in regular CCSNe. The energy density is much lower, and temperature do not reach the high values where neutrino cooling is important. Jets are likely to be the main energy transport from the CCSN to the ISM.
\newline
(3) If a fraction $\eta$ of the rest mass of the newly born CCSN is taken by jets, the explosion energy is
$E_{\rm CCSN} \approx 10^{57} (\eta/0.01)(M_{\rm SMBH}/10^5 M_\odot) \erg$.
\newline
(4) The presence of an accretion disk that launches jets around a supermassive star is reasonable.  We know that massive stars can harbor a large disk, e.g., \cite{Johnstonetal2015} found a disk of mass $\approx 10 M_\odot$ and of a size of $\approx 900 \AU$ around a young O star. Such massive disks are likely to launch bipolar jets/wind.
\newline
(5) The accretion from the ISM onto the SMYO is likely to operate as the SMYO collapses. This implies that the collapsing gas has a high value of specific angular momentum, an accretion disk with a constant axis is formed, and the jets launched by the newly born SMBH have a constant directions. In such a case only a small fraction of the star is expected to be blown away by the jets \cite{Gilkisetal2016}.
\newline
(5) Whalen et al. (2013) \cite{Whalenetal2013} claim that SMYO might undergo thermonuclear explosion. Such a process will prevent the YUSA scenario from taking place. Here I assume that sufficient number of MYSO do not experience thermonuclear explosion. Most important, I argue that even if thermonuclear explosion does not take place, a supernova explosion powered by jets does take place.

\section{Summary}
\label{sec:summary}

In Table \ref{table:compare} I listed eight types of astrophysical objects where jets, and more particularly the jet feedback mechanism (JFM), might operate (for more detail see \cite{Soker2016}).
In four of these classes jets are known to play significant, or even crucial, roles: in cooling flows, during galaxy formation, in YSO, and in planetary nebulae.
In CCSNe, CEE, GEE, and ILOTs the suggestion that a JFM takes place is still controversial. In particular in CCSNe.
\emph{I call for a refresh thinking and more detail studies of the possibility that jets play a large role in exploding massive stars and in the CEE.} If jets play a role in the CEE, then the GEE is likely to take place in some cases, as well as the operation of jets in many ILOTs.

In the present paper I presented a new speculative scenario where the first AGN were preceded by jets activity during the life of the supermassive young stellar object progenitor of the AGN.
A short and energetic phase of supernova took place between the supermassive YSO and the AGN phases.
I summarized the properties of this newly proposed Supermassive YSO~$\rightarrow$~CCSN~$\rightarrow$~AGN (YOSA) JFM scenario in Table \ref{table:YOSA}.
\emph{I further speculated that a JFM operating during the phase of the supermassive YSO started to establish the correlations between the mass of the SMBH and some properties of the stellar component of galaxies before the formation of the SMBH. }

I thank Elisabete M. de Gouveia Dal Pino for helpful comments.

\end{document}